# Metagenomic analysis reveals shared and distinguishing features in horse and donkey gut microbiome and maternal resemblance of the microbiota in hybrid equids

Yihang Zhou[1][†]

[1]*School of Life Sciences and Technology, Tongji University, Shanghai, China*

[†]co-corresponding author:

Yihang Zhou
E-mail: youngjoe@tongji.edu.cn

Keywords: horse, donkey, hinny, mule, gut microbiome, heterosis.

**Abstract**

Mammalian gut microbiomes are essential for host functions like digestion, immunity, and nutrient utilization. This study examines the gut microbiome of horses, donkeys, and their hybrids, mules and hinnies, to explore the role of microbiomes in hybrid vigor. We performed whole-genome sequencing on rectal microbiota from 18 equids, generating detailed microbiome assemblies. Our analysis revealed significant differences between horse and donkey microbiomes, with hybrids showing a pronounced maternal resemblance. Notably, Firmicutes were more abundant in the horse-maternal group, while Fibrobacteres were richer in the donkey-maternal group, indicating distinct digestive processes. Functional annotations indicated metabolic differences, such as protein synthesis in horses and energy metabolism in donkeys. Machine learning predictions of probiotic species highlighted potential health benefits for each maternal group. This study provides a high-resolution view of the equid gut microbiome, revealing significant taxonomic and metabolic differences influenced by maternal lineage, and offers insights into microbial contributions to hybrid vigor.

## Introduction

With co-evolution for millions of years, mammals and their gut microbes have established complex and crucial micro-ecological systems to support the optimal phylogeny [1]. Strong shreds of evidence have indicated the involvement of intestinal microbiota in digestive tract development, host immunity, nutrients utilization, etc [2-6]. Many great kinds of research on the microbiome in animals, such as humans [7-9], mice [10, 11], dogs [12, 13], cattle [14, 15], etc, have been deployed. The non-ruminant herbivore relies heavily on the activity of the colon and cecum as fermentation chambers for the hosts' life support [16]. For example, 65% of the horses' energy generation is accounted for by the short-chain fatty acids produced via fibrolytic bacteria [17].

One leading cause of horses' morbidity or mortality is gastrointestinal disease, which is ascribed to alterations in the gastrointestinal microbiota in the broad spectrum of clinics [16, 18-20]. The physiology of horses is also affected by the gut microbiota [21]. The use of probiotics in equine gastrointestinal disease is under wide research [22]. Despite the importance of the gut microbiome in horses, former research mainly focused on hindgut fermentation [23-26]. The fecal microbiota transits during the growth of horses [27]. Some researchers performed 16s or 18s or cetyltrimethylammonium bromide sequencing on horse's fecal or gastrointestinal bacteria communities [28-47], but the technology limitation disabled further understanding of microbial at the species level or a comprehensive scale. We still do not know the composition of horses' gut microbiota at high resolution.

Our knowledge gap of the microbiota of another non-ruminant herbivore, the donkey, is even larger. The donkey industry has great economic value due to the medical effect of donkey-hide gelatin, the similarity between donkey milk and human milk, the physical power of farming donkeys, etc [48-50]. Though recent research has studied the composition of donkey fecal microbiome [51-53], microbial functions of digestive tract digesta [54], biogeography, and functions of donkeys' microbiome [55], we still have no concept about the donkey's gut microbiome at species resolution, because all of these studies are based on 16s sequencing.

Heterosis, or hybrid vigor, is a well-known phenotype for more than 100 years since Charles Darwin's perception of many vegetables and Edward M East and George H Shull's rediscovery of maize [56]. However, the mechanisms of heterosis remain unclear. It is believed that genetic [57] and epigenetics [58] factors have significant roles. Recently, Maggie et. al. added a further dimension of interactions with the microbiome to the phenomenon [59-61]. It is logical to infer, will there be an effect of the microbiome in animals' heterosis, since we have identified the microbial role in plants' heterosis? However, few researchers studied the microbiome of heterosis. Edwards et. al. performed 16s sequencing on equine hybrids' fecal microbiota, [62, 63], but this sequencing tech only returned thousands of operational taxonomic units (OTUs).

In these cases, it is necessary to give a comprehensive and high-resolution understanding of the gut microbiome of non-ruminant herbivores and to provide insight and high-quality data for further animal heterosis-microbiome interaction studies. We performed whole-genome shotgun (WGS) metagenomic sequencing on the fecal microbiota of 6 horses, 6 donkeys, 3 hinnies, and 3 mules. We constructed comprehensive microbiome assemblies, generated detailed rectum microbiota composition maps, identified unexpected variances in taxonomy and metabolism, and discovered maternal-resemblance of the hybrids in the rectum microbiome. Our work will contribute to the microbiome community in many aspects.

**Results**

**Comprehensive assemblies of equids rectum microbiome using WGS metagenomic data.**

We performed Whole-genome-sequencing (WGS) on 18 rectum microbial samples across 4 equid species, including 6 horses samples, 6 donkeys samples, 3 hinnies samples, and 3 mules samples (Figure 1A). In total, 41.76 Gbps of metagenomic DNA sequences were generated. Each of the 18 samples sequenced. During quality control, we removed 0.19% of host contamination, 2.21% of low-quality bases and adapters, and 0.05% of viral sequences. The survived 40.77 Gbps were then sent to the assembly pipeline (see Methods). We constructed 3 microbiome assemblies, including the horse-mule-donkey-hinny microbiome assembly (HMDH), the horse microbiome assembly, and the donkey microbiome assembly. In this article, we mainly used HMDH assembly for analysis, which consists of 33,816,527 contigs, ranging from 300 bp to 733,264 bp with an N50 of 823 bp. The rarefaction plot (Figure 1B) showed the high quality of the HMDH assembly. We annotated the HMDH assembly at phylum, class, order, family, genus, and species levels (Figure 1C).

**Diversity analysis revealed the maternal resemblance in the rectum microbiome.**

We calculated the alpha diversity of the rectum microbiome in 4 species and observed a significant difference between horses and donkeys (Figure 1D). From the box plot, the median alpha-diversity of the mule is close to the horse, while the median alpha-diversity of the hinny is approaching the donkey. We then combined the horse and mule samples to be the horse-maternal group, the hinny and donkey samples to be the donkey-maternal group. The two-sided Wilcoxon signed-rank test returned a smaller $P$-value, which reduced from 0.026 (between horse and donkey) to 0.014 (between the horse-maternal group and donkey-maternal group) (Figure 1E). We then applied Principal coordinates analysis (PCoA) with the beta-diversity among the 4 species. Two ellipses separated the 18 samples into the horse-maternal group and donkey-maternal group with a significant $P$-value of 0.0002 (Figure 1F). Two principal coordinates explained more than 50% of the variance along the axis, which supported the reliability of the beta-diversity analysis. Both the alpha- and beta-diversity analyses revealed the maternal resemblance in the equids rectum microbiome.

**Comparison between the horse-maternal group and donkey-maternal group at**

**phylum, family, and genus level.**

We identified the abundant phylum of rectum microbial in 4 equids species, such as Firmicutes, Bacteroidetes, Proteobacteria, Verrucomicrobia, Actinobacteria, and Firbrobacteres, etc (Figure 1C). Interestingly, the Firmicutes in the horse-maternal group is much richer than it is in the donkey-maternal group. The Firbrobacteres, which is a relatively small phylum in most other species' microbiome, consists of an average of 6.8% relative frequency in the donkey-maternal group, compared to an average of 2.5% in the horse-maternal group (Figure 2A). The Firmicutes are known to play an important role in the digestion of carbohydrates, and the Firbrobacteres are important for the degradation of plant-based cellulose. The significant variance in the two phyla suggests different digestion circumstances between horses and donkeys, while similarity of digestion within maternal groups.

We then identified the featured family and genus between the horse-maternal group and donkey-maternal group (Figure 2B-C). The family Bacteroidaceae and its corresponding genus Bacteroides are relatively abundant in the donkey-maternal group and are the most featured taxa that distinguish the donkey-maternal group from the horse-maternal group. The number of featured taxa between the horse-maternal group and donkey-maternal group is imbalanced (5/1 at family level, and 13/1 at genus level, horse-maternal group v.s. donkey-maternal group). On the other hand, 15 of the top 30 abundant genera in rectum microbiota had a significant difference in relative frequency between the horse-maternal group and donkey-maternal group (Figure 2D). Only 30% of the top 30 abundant genera are richer in the donkey-maternal group, which is skewed from the expected 50% balance (Figure 2D). Considering the phylogenic short distance between horse and donkey, such a big variance in rectum microbiome was not expected. The imbalance in the number of featured taxonomy and the variance in the top abundant genus, as well as the difference in alpha-diversity, together suggest that the horse rectum microbiome has a higher diversity or complexity in taxonomy compared to the donkey. More interestingly, this taxonomy imbalance is maternal.

**The analysis of the rectum microbiome at the species level confirmed the maternal resemblance and variance between the two maternal groups.**

We counted the relative frequency of the top 30 abundant species in the rectum microbiome of 18 samples (Figure 3A). The horizontal cluster automatically separated the 18 samples into the horse-maternal group and donkey-maternal group. Vertically, the 30 microbial species were clustered into 3 blocks. In general, in the first block, the *Firmicutes bacterium CAG:555* and *Firmicutes bacterium CAG:110* are more abundant in the horse-maternal group, while the *Fibrobacter sp. UWP2* and *Lachnospiraceae bacterium G41* in the second block are richer in the donkey-maternal group. All 7 microbial species in the third block are more frequent in the donkey-maternal group. We then identified the top featured microbial species between the horse-maternal group and donkey maternal group (Figure 3B). The imbalance of species number was observed again (2/7, horse-maternal group v.s. donkey-maternal group).

Probiotics are well known to benefit host health through microbiome metabolism. We took advantage of machine learning to predict the probiotics species in 18 samples and counted their relative frequency (Figure 3C, see Methods). The two-sided Wilcoxon signed-rank test showed the marginal difference between the horse-maternal group and donkey-maternal group.

The analysis of rectum microbiome at species level confirmed both the maternal-resemblance, as well as the abundance variance in higher taxonomic level revealed in the previous analysis.

**Function and metabolism analysis indicated maternal resemblance and difference between the two maternal groups.**

We annotated the high-quality microbial reads with 5 independent databases, including the KEGG database (Kyoto Encyclopedia of Genes and Genomes), EggNOG database (v4.5), Pfam database (v35.0), the ENZYME database, and the MetaCyc database. The two-sided Wilcoxon signed-rank tests were applied to 18 samples across 5 databases, and the *P*-value matrixes were constructed (Figure 4A). Expectedly, the intragroup P-values are close to 1, while the intergroup P-values are close to 0, which indicated the maternal resemblance in metabolism.

We then identified the featured functions and pathways between the two maternal groups using the linear discriminant analysis (Figure 4B). From the EggNOG database, the horse-maternal group was featured in protein synthesis while the donkey-maternal group was featured in energy-related metabolism. From the KEGG database, the horse-maternal group was featured in transcription-related proteins, while the donkey-maternal group was featured in phosphate synthesis-related proteins.

**Methods**

**Quality control and preprocessing of metagenomic reads**

We generated 2.32 billion base pairs of DNA sequences per sample for each of the 18 rectum metagenomes. We used FastQC (v0.11.9) [64] to check the quality of reads. The reads were then trimmed to remove adapters and low-quality bases using Trimmomatic (v0.36) [65]. The passed reads from horses were aligned to the horse reference genome (GenBank: GCA_002863925.1). The passed reads from donkeys were aligned to the donkey reference genome (GenBank: GCA_003033725.1). The passed reads from mule and hinny were aligned to both horse and donkey reference genomes. This alignment step was to remove host contaminations. Then the survived reads were mapped to a viral genome assembled by National Center for Biotechnology Information (NCBI) to erase viral contaminations [66]. We used Burrows-Wheeler Aligner (BWA) (v0.7.17-r1188) for mapping [67], SAMtools (v1.6) [68, 69] and BEDTools (v2.25.0) [70] for bam format files processing.

**The rectum metagenome assembly, microbial taxonomy, and gene annotation**

The high-quality reads that survived from the previous quality-control pipeline were used as input in this section. We used PEAR (v0.9.11) [71] to join the paired-end (PE) reads to obtain longer single-end (SE) reads. We then used MEGAHIT (v1.1.2) to assemble the longer SE reads and un-joined PE reads into draft metagenome assembly. From the draft assembly, the redundant contigs were removed using CD-HIT (v4.7) [72] with a 97% identity threshold, and the contigs with sequence lengths less than 300 bases were deleted as noise using SeqKit (v2.1.0) [73]. To assign the taxonomy at phylum, class, order, family, genus, and species level, we used Kaiju (v1.7.3) [59] with its nr_euk database. To annotate the microbial genes in the metagenome assembly, we used the meta-gene predictor MetaGeneMark (v3.38) [74, 75].

**The taxonomy relative frequency analysis**

We mapped the high-quality reads back to the annotated metagenome assembly can calculate the relative frequency of each taxonomy at phylum, class, order, family, genus, and species level for all 18 samples. We then applied two-sided Wilcoxon signed-rank tests using R package ggpubr (v0.4.0), pheatmap (v1.0.12) and grid (v4.0.5), plotted heatmaps using R package ggplot2 (v 3.3.5) [76] and reshape2 (v4.0.5) [77], calculated qvalue using the R package qvalue (v2.22.0). Some processing scripts were written using Python (v3.7.7).

**The diversity analysis**

The alpha-diversity was performed with the Simpson index [78]. The beta-diversity was analyzed based on the Bray-Curtis dissimilarity [79]. Both indices were calculated using R package vegan v2.5.7 [13]. The Principal Coordinates Analysis (PCoA) was plotted using ggplot2 (v 3.3.5) [76, 80].

**Identification of the most discriminative taxonomy features**

With the relative frequency of taxonomy at phylum, class, order, family, genus, and species level, we performed the linear discriminant analysis Effect Size (LEfSe v1.1.1) [81] analysis to identify the most feature taxonomy between the horse-maternal group and donkey-maternal group.

**Prediction of the probiotic**

We remove the contigs in metagenome assembly with a relative frequency of less than 0.001% to remain 2,830 abundant contigs. We predicted the probability of these contigs being probiotics or non-probiotics with the support-vector machine using iProbiotics [82]. We took contigs with probability larger than 80% as probiotics, less than 20% as non-probiotics, and between 20%-80% as unknown. The high-quality reads were mapped to these contigs to calculate the relative frequency.

**Function and metabolism analysis**

We used HUMAnN3 [83] to profile the function and metabolism pathways enrichment of the 18 samples. The high-quality reads were annotated using KEGG database (Kyoto Encyclopedia of Genes and Genomes) [84], EggNOG database (v4.5) [85], Pfam database (v35.0) [86], the ENZYME database [87], and the MetaCyc database [88]. The reference database we used was "UniRef90" and the abundances of each gene family were normalized in counts per million reads (CPM) units. For each database, we applied the two-sided Wilcoxon signed-rank test among the horse, mule, donkey, hinny, horse-maternal group, and donkey-maternal group. The $P$-values of the comparison were visualized in the heatmap. The LEfSe analysis was also performed between horse-maternal group and donkey-maternal group to identify featured pathways from 5 independent databases.

**Discussion**

This study offers an in-depth analysis of the rectal microbiome among different equids—horses, donkeys, hinnies, and mules—through the use of whole-genome sequencing metagenomics. The results highlight the intricate relationships between microbial diversity, composition, and function in these non-ruminant herbivores, emphasizing the significant effects of maternal lineage and hybrid vigor.

The microbial communities in horses and donkeys were found to be distinctly different, correlating with their unique digestive strategies and dietary adaptations. Horses and their hybrids showed a dominance of Firmicutes, bacteria known for fermenting carbohydrates, while donkeys and their hybrids had higher levels of Fibrobacteres, essential for breaking down cellulose. These variations likely represent evolutionary adaptations to their specific dietary needs.

Functionally, there were significant differences between the microbial communities of the two maternal lineages. The horse-maternal group demonstrated a higher capacity for protein synthesis, while the donkey-maternal group was enriched with genes associated with energy metabolism. These functional differences reflect the diverse metabolic demands imposed by the different dietary habits of these species, indicating a metabolic optimization aligned with their nutritional environments.

Utilizing machine learning techniques, our study also identifies potential probiotic candidates unique to each maternal lineage. This finding opens up new possibilities for targeted microbial interventions to improve health and productivity in equids. Given the substantial impact of gastrointestinal diseases on these animals' morbidity and mortality, such probiotic applications could revolutionize equine healthcare.

Moreover, the strong maternal influence observed in the microbiomes of hybrids supports the idea that maternal lineage is a crucial factor in shaping microbial communities. This finding supports the concept of heterosis, where hybrids exhibit enhanced traits possibly driven by microbial diversity. This novel insight into the microbiome's role in hybrid vigor suggests that further research is needed to understand how microbial diversity interacts with genetic and epigenetic factors to promote heterosis.

Future research should focus on exploring the dynamic interactions between the host genome, epigenome, and microbiome. Such studies could elucidate the various factors that shape the microbiome and its role in expressing heterosis. Longitudinal studies tracking microbial community evolution from birth to maturity would provide valuable insights into the stability and adaptive changes within these communities. Additionally, understanding how these microbiomes respond to dietary and environmental changes will further clarify their role in the health and adaptation of equids.

In summary, this research significantly enhances our understanding of the equid gut microbiome, highlighting critical taxonomic and metabolic differences influenced by maternal lineage. These findings underscore the microbiome's vital role in animal physiology, health, and the expression of hybrid vigor. This study lays the groundwork for future research into the complex interactions between genetics, microbiomes, and phenotypic outcomes in non-ruminant herbivores.

**Figure 1. Deep whole-genome-sequencing in horse, mule, donkey, and hinny rectum microbiota revealed maternal resemblance.**

(**A**) Representative illustration of the species pedigree. Maternal horse (*orange*) and paternal donkey (*gray*) cross breed to mule (*yellow*). Maternal donkey and paternal horse cross breed to hinny (*blue*).
(**B**) Refraction plot of the number of identified bacterial species per sample.
(**C**) Relative microbial frequency in the rectum microbiota at the phylum level for the 6 horse, 3 mules, 3 hinnies, and 6 donkey samples.
(**D**) Boxplots of alpha diversity of horse, mule, hinny, and donkey observed at the species level measured using the Simpson index. Statistical significance was assessed by a two-sided Wilcoxon signed-rank test.
(**E**) Box plots of alpha diversity of the horse-maternal group and the donkey-maternal group observed at the species level measured using the Simpson index. Statistical significance was assessed by a two-sided Wilcoxon signed-rank test.
(**F**) The Principal Coordinates Analysis (PCoA) plots of beta diversity among horse, mule, donkey, and hinny rectum microbiota using Bray-Curtis distance. Ellipse (level = 0.95) and statistical significance (two-sided Wilcoxon signed-rank test) were processed between the horse-maternal group and donkey-maternal group.

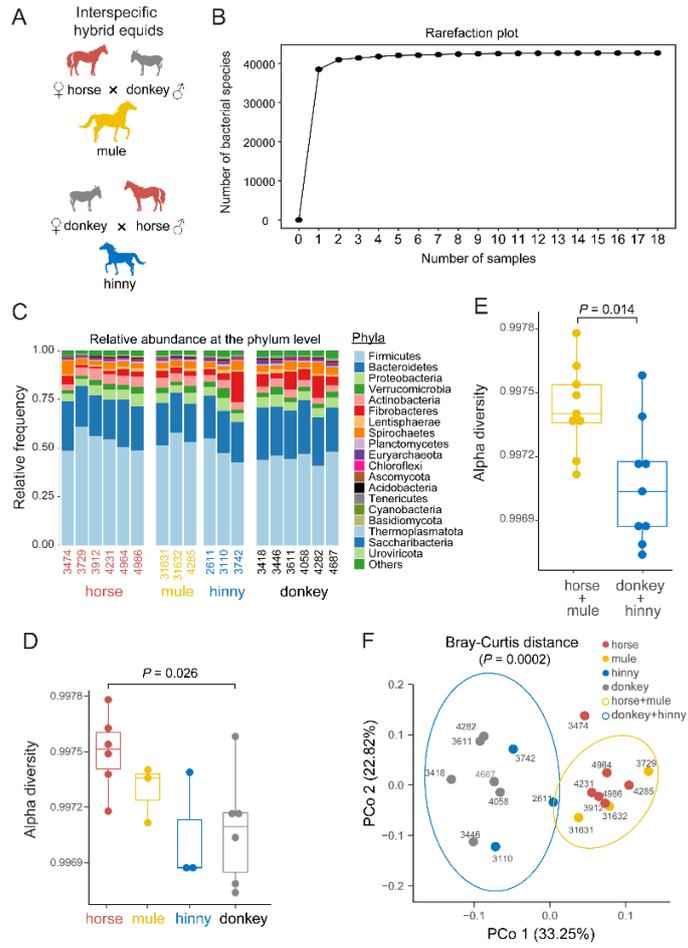

**Figure 2. Taxonomy comparison between the horse-maternal group and donkey-maternal group at phylum, family, and genus level.**
(**A**) Boxplots of relative microbial frequency for two top abundant phyla (Firmicutes and Fibrobacteres) showing significant difference between horse-maternal group and donkey-maternal group (two-sided Wilcoxon signed-rank test).
(**B-C**) The most featured families (**B**) and genus (**C**) with linear discriminant analysis (LDA) score ⩾3 in the horse-maternal group and donkey-maternal group.
(**D**) Boxplots of relative microbial frequency (log) of top 30 genera in rectum microbiota. Star-sign shows a significant difference between the horse-maternal group (*yellow*) and donkey-maternal group (*blue*) (two-sided Wilcoxon signed-rank test). The genus is ranked descending along the x-axis (*yellow*: on the average horse-maternal group has higher abundance, *blue*: on the average donkey-maternal group has higher abundance).

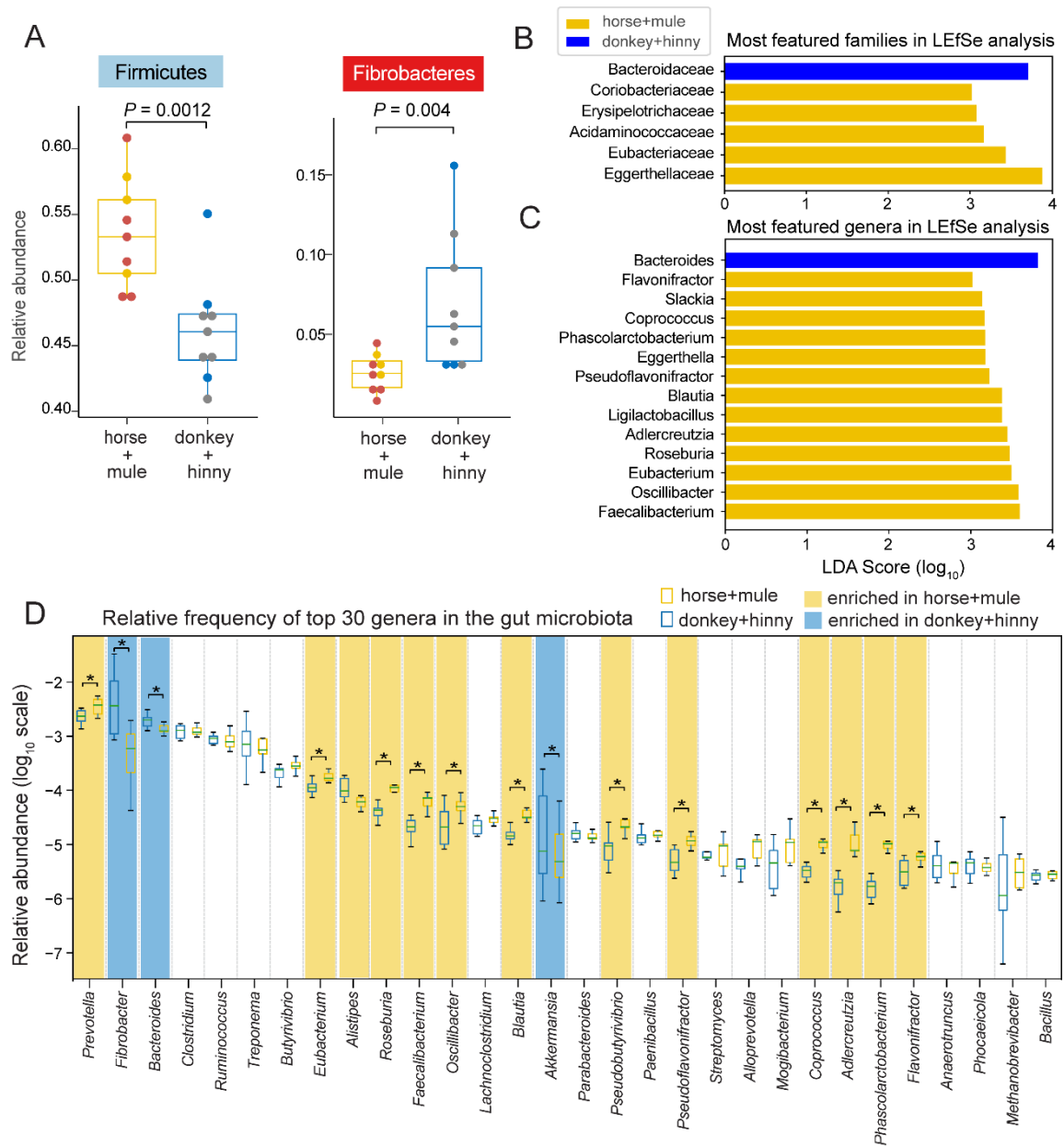

**Figure 3. Taxonomy analysis between the horse-maternal group and donkey-maternal group at the species level.**

(**A**) Cluster heatmaps of the relative frequency for the top 30 abundant species in equids rectum microbiota. Horse-maternal group and the donkey-maternal group were automatically clustered apart. The sample IDs listed on the right were colored.

(**B**) The most featured species with linear discriminant analysis (LDA) score ⩾3 in the horse-maternal group and donkey-maternal group.

(**C**) Boxplots of relative microbial frequency of predicted probiotics species in rectum microbiota in horse, mule, hinny, and donkey. Statistics show a marginal difference between the horse-maternal group (*yellow*) and donkey-maternal group (*blue*) (two-sided Wilcoxon signed-rank test). The prediction was based on a support-vector machine from iProbiotics[??].

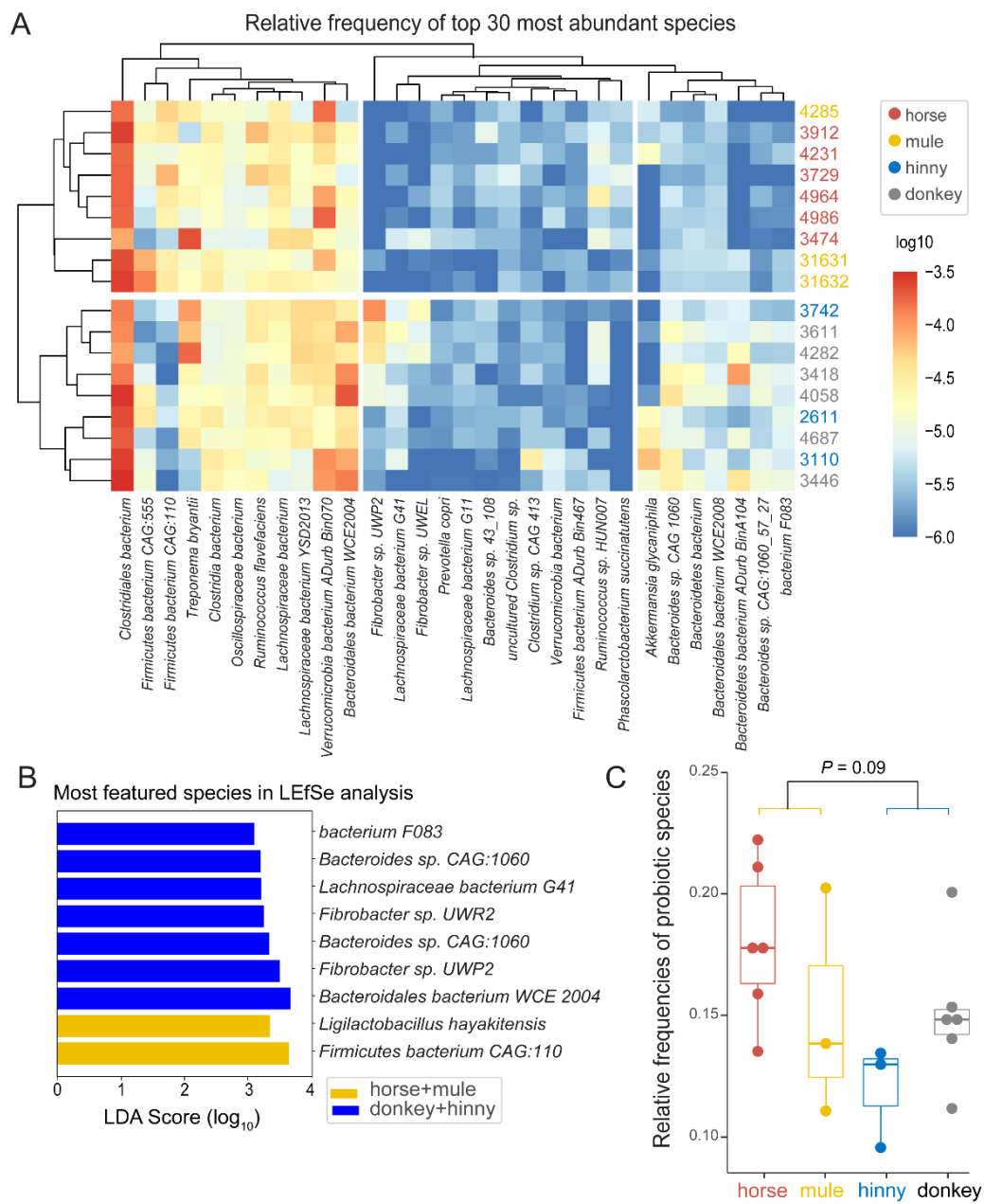

**Figure 4. Function and metabolism analysis between the horse-maternal group and donkey-maternal group from 5 aspects.**

(**A**) Heatmaps of the P-value matrix show maternal resemblance at 5 function and metabolism aspects. The pink frame on the left is function analysis combining Enzyme Commission as upper matrix and MetaCyc as the lower matrix. The green frame on the right is metabolism analysis combining EggNog as the upper matrix and KEGG Ortholog as the lower matrix. Each row and column was labeled the corresponding species. The red side means a larger P-value, indicating higher similarity. The blue side means less P-value, indicating more difference. The calculation of P-value matrixes was described in the **Method** part.

(**B**) The top 10 most featured pathways or categories with linear discriminant analysis (LDA) score ≥3 in horse-maternal group and donkey-maternal group.

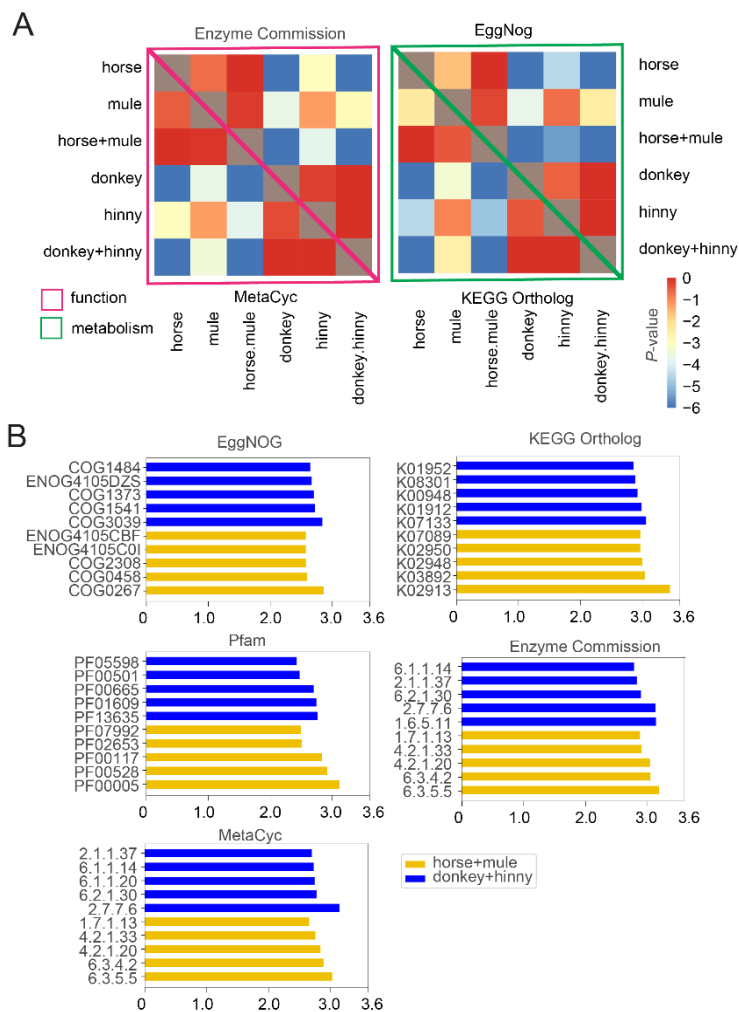